\documentclass{mem}
\usepackage{natbib}\usepackage{txfonts}\usepackage{balance}
\usepackage{graphicx}
\usepackage{txfonts}
\usepackage[a4paper]{hyperref}
\idline{79}{1}
\begin{document}

\def\teff{$T\rm_{eff }$}
\def\kms{$\mathrm {km s}^{-1}$}
\def\logtaustd{$\log\tau_{\rm{5000}}$}

\title{Abundance Analysis of the Halo Giant HD122563 with Three-Dimensional Model Stellar Atmospheres}

%\subtitle{}

\author{
R.\,Collet\inst{1},
{\AA}.\,Nordlund\inst{2},
M.\,Asplund\inst{1},
W.\,Hayek\inst{3,1}
\and R.\,Trampedach\inst{3}
}

%%\offprints{R. \,Collet, \email{remo@mpa-garching.mpg.de}}

\institute{Max-Planck-Institut f{\"u}r Astrophysik,
Postfach 1317, D--85741 Garching b. M{\"u}nchen, 
Germany; \email{remo@mpa-garching.mpg.de}
\and
Niels Bohr Institute, University of Copenhagen, Juliane Maries Vej 30, DK--2100, Copenhagen, Denmark
\and
Research School of Astronomy and Astrophysics,
Mount Stromlo Observatory,
Cotter Road,
Weston ACT 2611,
Australia}

\authorrunning{Collet et al.}

\titlerunning{3D Abundance Analysis of HD122563}

\abstract{We present a preliminary local thermodynamic equilibrium 
(LTE) abundance analysis of the template halo red giant HD122563 
based on a realistic, three-dimensional (3D), time-dependent, 
hydrodynamical model atmosphere of the very metal-poor star.
 We compare the results of the 3D analysis with the abundances 
derived by means of a standard LTE analysis based on a classical, 
1D, hydrostatic model atmosphere of the star.
Due to the different upper photospheric temperature stratifications 
predicted by 1D and 3D models, we find large, negative, 3D$-$1D LTE 
abundance differences for low-excitation OH and Fe~{\sc i} lines.
 We also find trends with lower excitation potential in the derived 
Fe LTE abundances from Fe~{\sc i} lines, in both the 1D and 3D analyses.
Such trends may be attributed to the neglected departures from LTE 
in the spectral line formation calculations.

}

\maketitle{}

\section{Introduction}
HD122563 is one of the most widely studied halo stars
and is generally considered as a template for metal-poor red giants
in abundance analyses \citep[e.g.,][]{barbuy03,cowan05,aoki07}.
Up to now, all spectroscopic analyses of this star have relied on the use
of classical, one-dimensional (1D), stationary, hydrostatic model stellar
atmospheres.
In this contribution, we present some preliminary results from
the first abundance analysis of HD122563 based on a three-dimensional,
time-dependent, hydrodynamical model atmosphere of the halo giant.

\section{Methods}
%% Figure Atmos
\begin{figure}[]
\resizebox{\hsize}{!}{\includegraphics{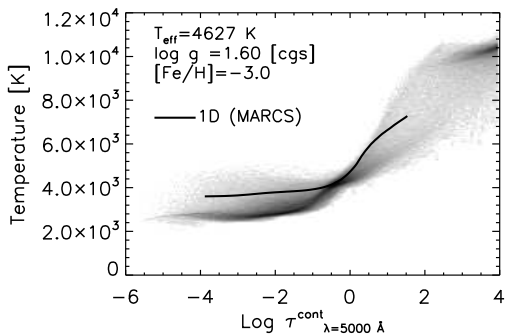}}
\caption{\footnotesize
\emph{Gray shaded area}: Temperature distribution as a function of 
continuum optical depth at $\lambda=5000$~{\AA} in the 3D metal-poor red 
giant surface convection simulation used in the present analysis of HD122563; 
darker areas indicate values with higher probability.
\emph{Solid line}: Temperature stratification in the 1D {\sc marcs} 
model atmosphere constructed for the same stellar parameters.}
\label{fig:atmos}
\end{figure}
We use the {\sc stagger-code} 
\citep{nordlund95}\footnote{\url{http://www.astro.ku.dk/~kg/Papers/} 
\url{MHD_code.ps.gz}}
to carry out a 3D, radiative, hydrodynamical, simulation of convection 
at the surface of a red giant with approximately the same stellar parameters as
HD122563,
that is with an effective temperature {\teff}${\approx}4600$~K,
a surface gravity ${\log}{g}=1.6$~(cgs), and a scaled solar chemical composition 
\citep{grevesse98} with [X/H]$=-3~$\footnote{[A/B]${\equiv} 
\log{(n_\mathrm{A}/n_\mathrm{B})}-\log{(n_\mathrm{A}/n_\mathrm{B})}_\odot$, 
where $n_\mathrm{A}$ and $n_\mathrm{B}$ are the number densities of elements 
A and B, respectively, and the circled dot refers to the Sun.} 
for all metals.
The equations for the conservation of mass, momentum, and energy 
are solved together with the radiative transfer equation on a discrete 
mesh with $480{\times}480{\times}240$ numerical resolution for a 
representative volume of stellar surface deep enough to cover about 
eleven pressure scale heights and large enough to incorporate about 
ten granules (${\sim}3700{\times}3700{\times}1100$~Mm$^3$).

We adopt open boundaries at the top and at the bottom of the 
simulation domain, and periodic boundary conditions at the sides.
We use a realistic equation-of-state \citep{mihalas88} and 
continuous and line opacities by \cite{gustafsson75} and 
\cite{kurucz92,kurucz93}.
To compute the radiative heating rates entering the energy conservation 
equation, we group the opacities in four \emph{opacity bins} 
\citep{nordlund82} and solve the radiative transfer equation
for each bin along the vertical and eight inclined rays.
We assume local thermodynamic equilibrium (LTE) and treat scattering 
as true absorption.

%% Figure Fe I and Fe II
\begin{figure*}[]
\resizebox{\hsize}{!}{
\includegraphics{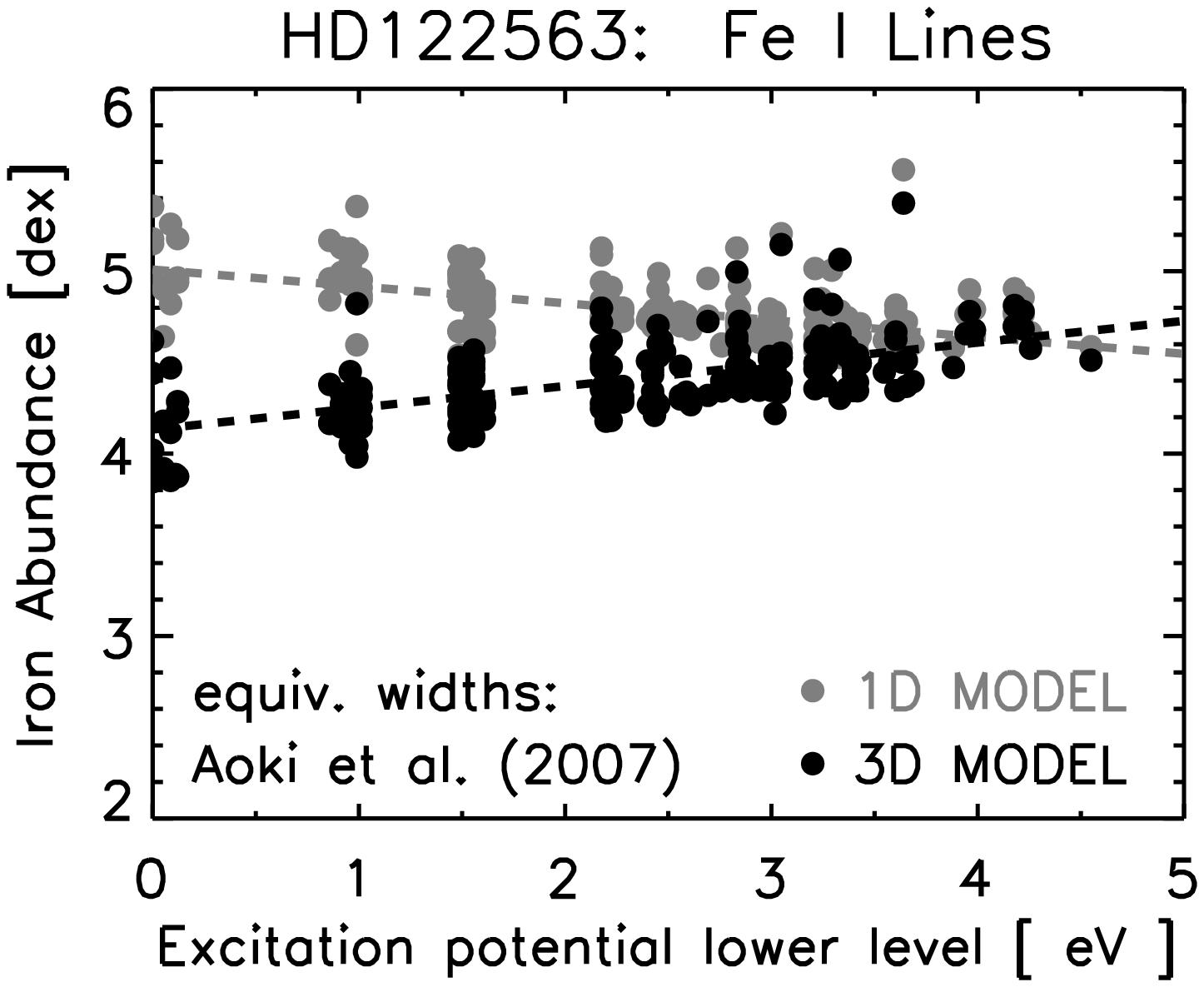}
\includegraphics{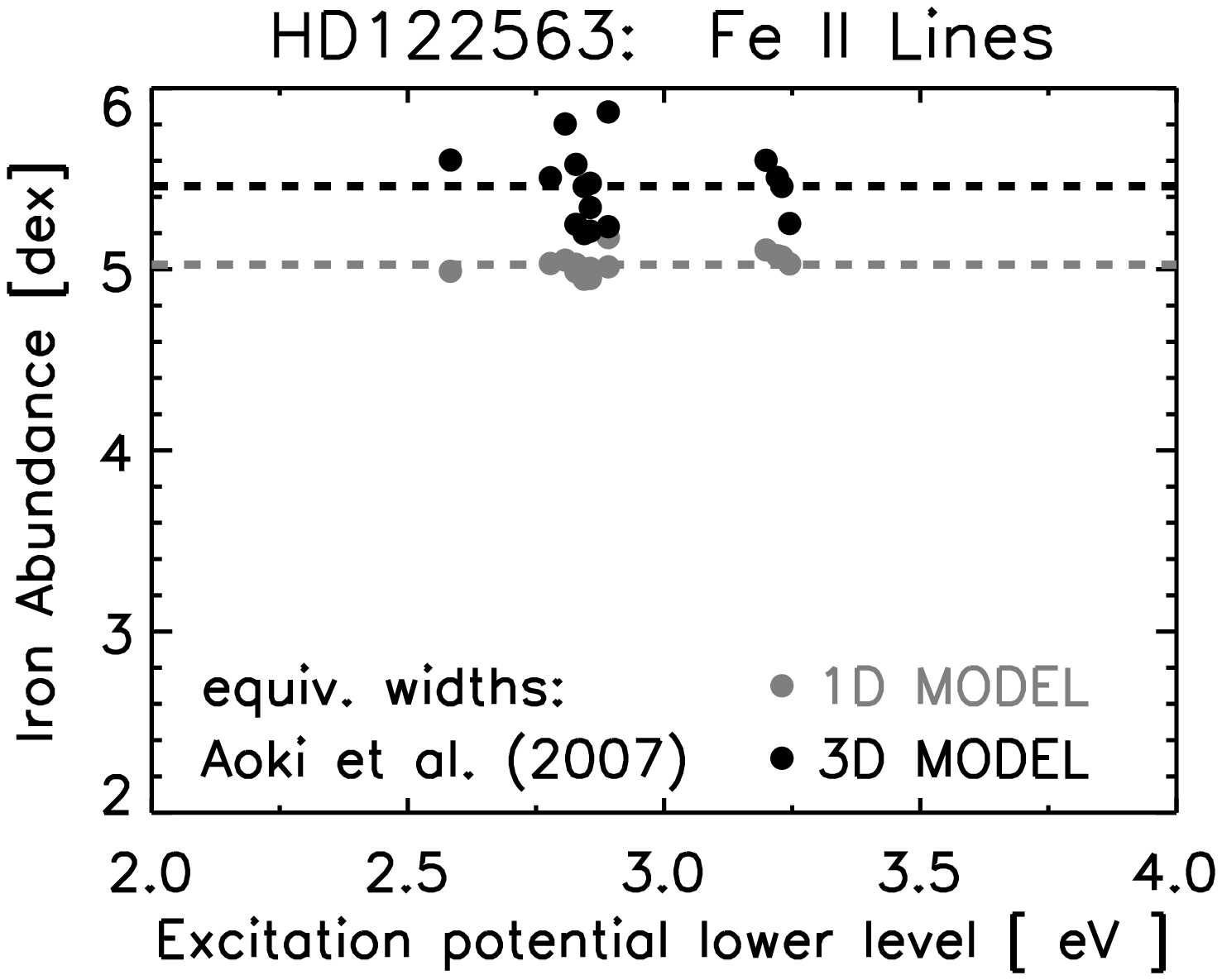}
}
\caption{\footnotesize
HD122563: Iron abundance derived from Fe~{\sc i} (\emph{left panel})
and Fe~{\sc ii} lines (\emph{right panel}) 
as a function of lower excitation potential; black (grey) filled 
symbols represent the results of the 3D (1D) analysis. 
The equivalent widths of the lines are taken from \cite{aoki07}. }
\label{fig:feabund}
\end{figure*}

The predicted temperature and density stratification 
as well as the dynamics of the gas flows are qualitatively 
very similar to the ones reported for previous simulations 
of convection at the surface of red giants \citep{collet07} and
solar-type stars \citep{stein98,asplund01}.
Warm, isentropic gas ascends from the bottom and, as it approaches 
the optical surface, it begins to cool by photon losses.
Near the surface, the opacity behaves as a strongly sensitive function of 
temperature and decreases rapidly as the gas cools radiatively. 
As the gas becomes less opaque, photons escape more easily, 
lowering the temperature even further and cooling the gas to the 
point that it becomes negatively buoyant and starts to sink downward. 
This positive feedback causes  a sudden drop of the gas temperature 
from ${\sim}10\,000$~K to ${\sim}5\,000$~K, which in the present 
metal-poor red giant simulation takes place within a distance of 
typically ${\sim}30$~Mm or less from the optical surface
resulting in very steep vertical temperature gradients.
The optical surface also appears to be very corrugated, with optical
depth unity occurring over a large range of geometrical 
depths (${\sim}250$~Mm).

The temperature stratification as a function of optical depth 
for a typical simulation snapshot is shown in figure~\ref{fig:atmos}.
Superimposed is the corresponding stratification from a 1D, LTE, 
plane-parallel, hydrostatic {\sc marcs} model atmosphere \citep{gustafsson75,asplund97}
computed for the same stellar parameters. 
The temperatures in the upper photospheric layers of the 3D simulation
are significantly lower (by ${\sim}1000$~K, on average) than in 1D.
This disparity can be explained considering that in stationary 
hydrostatic and in time-dependent hydrodynamical models
the mechanisms controlling the energy balance
in the upper photosphere are effectively different: radiative equilibrium, in 1D, 
and radiative heating and adiabatic cooling due to gas expansion,
in 3D \citep[see][]{asplund99,collet07}.

We use the red giant surface convection simulation as a 3D, 
time-dependent, hydrodynamical model atmosphere to compute spectral 
lines for neutral and singly ionized Fe as well as for a number
of molecules (e.g., OH, CH, and CN), under the assumption of LTE.
We derive Fe and O abundances by reproducing the measured equivalent 
widths of Fe~{\sc i} and Fe~{\sc ii} \citep{aoki07}
and infrared (IR) vibrational-rotational OH lines at $1.6$~$\mu$m 
\citep{barbuy03}.
We then compare the resulting abundances with the ones derived 
with the corresponding 1D {\sc marcs} model atmosphere, assuming LTE and adopting 
a micro-turbulence of $2.0$~{\kms} for the 1D spectral line formation calculations.

\section{Results and Discussion}
Figure~\ref{fig:feabund} shows the Fe abundance derived from 
Fe~{\sc i} and Fe~{\sc ii} spectral lines as a function of 
lower excitation potential.
The systematic cooler upper photospheric temperature
stratification of the 3D model atmosphere compared with the 1D model
result in significant differences between the 3D and 1D LTE 
ionization equilibria. In particular, the fraction of neutral Fe 
in those layers is, on average, higher in the 3D model than in 1D.
Hence, at a given abundance, synthetic Fe~{\sc i} lines tend to be 
stronger in 3D than in 1D, and a lower abundance is therefore required in 3D 
to match the observed equivalent widths of Fe~{\sc i} lines.
Conversely, the Fe abundance derived from Fe~{\sc ii} lines is
larger in the 3D case than in the 1D one.
As a consequence, the difference between the Fe abundance values
 determined from Fe~{\sc i} and Fe~{\sc ii} lines actually increases
 when going from 1D to 3D. 
We also observe trends in the Fe abundance derived from
Fe~{\sc i} lines with excitation potential in both 1D and 3D, 
but with opposite slopes in the two cases.
Such trends as well as the difference between the Fe abundance 
determinations from neutral and singly ionized iron lines are 
possibly due to the neglected departures from LTE in the line 
formation calculations and/or to  shortcomings still present 
in the model atmospheres.

%% Figure OH
\begin{figure}[]
\resizebox{\hsize}{!}{ \includegraphics{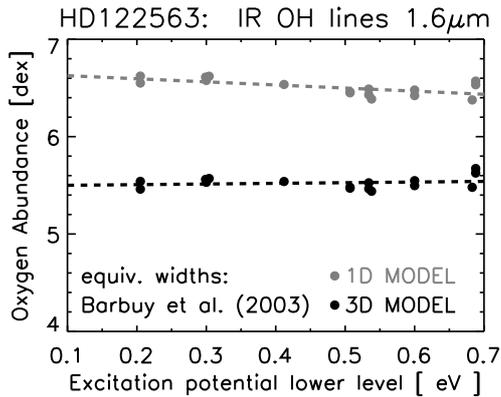} }
\caption{\footnotesize
HD122563: Oxygen abundance derived from IR OH lines at $1.6$~$\mu$m 
as a function of lower excitation potential; 
black (grey) filled symbols indicate the results of the 3D (1D) analysis. 
The equivalent widths of the lines are taken from \cite{barbuy03}. }
\label{fig:ohlines}
\end{figure}

Due to the highly non-linear temperature sensitivity of molecule
formation, the differences between the 3D and 1D temperature 
stratifications also lead to radically different molecular equilibria
in the two kinds of models.
From the analysis of CH lines at $4300$~{\AA}, the derived 3D carbon 
abundance is ${\sim}0.2$~dex \emph{lower} than the 1D value. 
The 3D$-$1D nitrogen abundance difference estimated from the analysis of CN lines at $3880$~{\AA}
is even more pronounced and  reaches ${\sim}1$~dex.
Figure~\ref{fig:ohlines} shows the 3D and 1D oxygen abundances 
determined from the IR OH lines: again, the 3D abundance is lower 
by ${\sim}1$~dex than the 1D one.
The 3D oxygen abundance values show no trend with lower excitation 
potential, which is reassuring as it suggests that the temperature
gradient in the OH line formation region is probably well reproduced by the 3D simulation.
The 1D oxygen abundance values show instead a decreasing,
although relatively shallow, trend with excitation potential.

We also determine the 3D and 1D oxygen abundance from the forbidden 
[O~{\sc i}] line at $6300$~{\AA}: we find that the 3D$-$1D abundance 
difference amounts to only $-0.15$~dex for this line. 
When we apply such 3D$-$1D abundance corrections to the oxygen abundances
derived by \cite{barbuy03} we find the corrected values from 
[O~{\sc i}] and OH lines to be discrepant.
The reasons for this discrepancy need be investigated further.
It may indicate that the temperature stratification predicted 
by the present simulations is still not accurate enough.
We are currently re-running the metal-poor red giant simulation with an 
improved and more accurate binning scheme by \cite{trampedach09}: 
preliminary tests show that the resulting vertical temperature gradient in the atmosphere
tends to become shallower and the differences with the 1D upper photospheric stratification
appear smaller. 
We need however to wait for the simulation to fully relax before 
drawing conclusions about the consequences for spectral line formation
and abundance analysis.
Also, one would need to investigate other sources of possible systematic
errors such as the treatment of scattering as true absorption in the convection
simulations, and, more importantly, the possibility of departures
from LTE in the calculation of excitation, ionization and molecular equilibria.

\bibliographystyle{aa}
\bibliography{collet}

\end{document}